\documentclass[journal]{IEEEtran}
\usepackage{graphicx}

\begin{document}
\thispagestyle{empty}

\title{An Optimization of the FPGA Based Wavelet \\Trigger in Radio Detection of Cosmic Rays}

\author{Zbigniew~Szadkowski,~\IEEEmembership{Member,~IEEE} for the Pierre Auger Collaboration% <-this % stops a space
\thanks{Manuscript received June 03, 2014.}  

\thanks{Zbigniew~Szadkowski is with the University of \L{}\'od\'{z}, Department of Physics and Applied Informatics, 
Faculty of High-Energy Astrophysics, 90-236 \L{}\'od\'{z}, Poland,
(e-mail: zszadkow @kfd2.phys.uni.lodz.pl, phone: +48 42 635 56 59).}%
}

\maketitle

\begin{abstract}

Experiments that observe coherent radio emission from extensive air showers induced by ultra-high energy
cosmic rays are designed for a detailed study of the development of the electromagnetic part of air showers.
Radio detectors can operate with 100\% up time as e.g. surface detectors based on water-Cherenkov tanks. They
are being developed for ground-based experiments (e.g. the Pierre Auger Observatory) as another type of air shower
detector in addition to the fluorescence detectors, which operate with only $\sim$10\% of duty in dark nights.
The radio signals from air showers are caused by the coherent emission due to geomagnetic radiation and charge excess
processes. Currently used self-triggers in radio detectors often generate a dense stream of data, which is
analyzed afterwards. Huge amounts of registered data requires a significant man-power for the off-line analysis.
An improvement of the trigger efficiency becomes a relevant factor. The wavelet trigger, which investigates on-line
a power of radio signals ($\mathrm{\sim V^2/R}$) is promising with respect to current designs.

In this work, Morlet wavelets with various scaling factors were used for an analysis of real data from the Auger
Engineering Radio Array (AERA) and for an optimization of the utilization of the resources in an FPGA. The wavelet
analysis showed that the power of events is concentrated mostly in a limited range of the frequency spectrum
(consistent with a range imposed by the input analog band-pass filter). However, we found several events with
suspicious spectral characteristics, where the signal power is spread over the full band-width sampled by a
200 MHz digitizer with significant contribution of very high and very low frequencies. These events may not origin
from cosmic ray showers but can be human-made contaminations. The engine of the wavelet analysis can be
implemented into the modern powerful FPGA and can remove suspicious events on-line to reduce the trigger rate.

\end{abstract}


\IEEEpeerreviewmaketitle

\section{Introduction}

\IEEEPARstart{R}{esults} from various cosmic rays experiments located 
on the ground level,
point to the need for very large aperture detection systems
for ultra-high energy cosmic rays. With its nearly 100\% duty
cycle, its high angular resolution, and its sensitivity to the longitudinal
air-shower evolution, the radio technique is particularly 
well-suited for detection of Ultra High-Energy Cosmic Rays 
(UHECRs) in large-scale arrays.
The present challenges are to understand the emission mechanisms
and the features of the radio signal, and to develop an
adequate measuring instrument. Electron-positron pairs generated
in the shower development are separated and deflected by
the Earth magnetic field, hence introduce an electromagnetic
emission \cite{Alan}\cite{Falcke_Gorham}. During shower development, charged particles
are concentrated in a shower disk of a few meters thickness.
This results in a coherent radio emission up to about 100 MHz. Short
but coherent radio pulses of 10 ns up to a few 100 ns duration
are generated with an electric field strength increasing approximately
linearly with the energy of the primary cosmic particle
inducing the extended air showers (EAS), i.e. a quadratic dependence
of the radio pulse energy vs. primary particle energy.
In contrast to the fluorescence technique with a
duty cycle of about 12\% (fluorescence detectors can work only during 
moonless nights), the radio technique allows nearly full-time measurements
and long range observations due to the high transparency
of the air to radio signals in the investigated frequency range. 

The radio detection technique will be complementary to the
water Cherenkov detectors and allows a more precise study of
the electromagnetic part of air showers in the atmosphere. In addition
to a strong physics motivation, many technical aspects relating
to the efficiency, saturation effects and dynamic range, the
precision for timing, the stability of the hardware developed, deployed
and used, as well as the data collecting and system-health
monitoring processes will be studied and optimized. 

One of the currently developed technique is an radio signals 
power estimation based on the wavelet transforms.
The main motivation of a development based on much more sophisticated 
algorithms is that the efficiency of the radio self-trigger is often very low. 
Mostly registered events contain only noise. The significant improvement 
of the trigger efficiency is the crucial factor. A lot of data analyzing 
off-line requires non-negligible amount of man-power. Much wiser
approach is much more efficient trigger. The presented wavelet trigger 
is an alternative proposal to the currently operating algorithms. 

\section{Timing and Limitations}

The radio signal is spread in time interval of an order of couple hundred nanoseconds, 
most of registered samples gave a time interval below 150 ns.
The frequency window for the maximal antenna efficiency 
is $\sim$30 - 80 MHz. This range is additionally filtered by a band-pass filter. According to the Nyquist
theorem the sampling frequency should be twice higher than the maximal 
frequency in a investigated spectrum. The anti-aliasing filter should 
have the cut-off frequency of $\sim$85 MHz.
Taking into account some width of the transition range for the filter 
(from pass-band to stop-band)
the final sampling frequency should not be lower than 170 MHz (200 MHz in our considerations).
This frequency corresponds to 5 ns between rising edges of the clock.
The widths of the radio pulses $\sim$100 ns are consistent with the theoretical 
estimations \cite{Falcke}, \cite{Huege} and the LOPES data 
\cite{Petrovic}, \cite{LOPES}. 
The width of $\sim$100 ns by the 200 MSps sampling corresponds to 20 time bins.
A minimal length of the FFT routine covering the pulse width equals 32.

Only a single FFT32 routine for on-line calculation of Fourier coefficients 
for data is needed. Fourier coefficients for various wavelets
can be calculated earlier and be available for final the power estimation as constants.
Nevertheless, we also consider 16-point FFT routine to minimize a FPGA resource occupancy
and to increase a safety margin for a registered performance.
The Quartus$\textsuperscript{\textregistered}$ II environment for 
the Altera$\textsuperscript{\textregistered}$ FPGA programming provides
parameterized FFT routines with various architectures: streaming, variable streaming, burst and buffered burst.
For the variable streaming provides also fixed or floating-point 
algorithms with natural 
or bit-reversed order. However, all routines deliver the FFT 
coefficients in a serial form.
No any Altera$\textsuperscript{\textregistered}$ routine allows calculating 
all FFT coefficients simultaneously.

\section{Wavelets}

The wavelet transform can be used to analyze time
series that contain non-stationary power at many different
frequencies \cite{Daubechies}.
Let us investigate a time series X, with values of $x_n$, at time index n. 
Each value is separated in time by a constant time interval $\Delta$t. 
The wavelet transform $W_n(s)$ is just the inner product (or convolution) 
of the wavelet function $\psi$ with our original time series: 
\begin{equation}
W_n(s) = \sum\limits_{m=0}^{N-1} x_m \psi^{*}  \left[\frac{(m-n)\Delta t}{s}\right]
\label{wavelet}
\end{equation}
where s denotes the frequency scale and the asterisk (*) complex conjugate, respectively. 
A variation of the wavelet scale s and a transition along the
localized time index n allow a construction of a dependence
showing both the amplitude of any features vs. the
scale and how this amplitude varies with time. 
Although it is possible to calculate the wavelet transform
according to (\ref{wavelet}), it is considerably faster to do the calculations
in the Fourier space.

To approximate the continuous wavelet transform,
the convolution (\ref{wavelet}) should be done N times for each
scale, where N is the number of points in the time series. 
However, the choice of doing all N convolutions is arbitrary, 
we can decrease N to a smaller
number to reduce computation time if the precision is satisfactory. 
By choosing N points, the convolution theorem allows 
performing all N convolutions simultaneously in the Fourier space
using the discrete Fourier transform (DFT). The DFT
of $x_n$ is
\begin{equation}
X_k = \frac{1}{N} \sum\limits_{n=0}^{N-1} x_n e^{-2\pi k n/N}
\label{Fourier}
\end{equation}
where k = 0,...,N - 1 is the frequency index. In the
continuous limit, the Fourier transform of a function
$\psi(t/s)$ is given by $\Psi(s\omega)$. By the convolution theorem,
the wavelet transform is the inverse Fourier transform
of the product:
\begin{equation}
W_n(s) = \sum\limits_{k=0}^{N-1} \bar{X}_k \bar{\Psi}^{*}(s \omega_k) e^{i \omega_k n \Delta t }
\label{wavelet_freq}
\end{equation}
where the angular frequency is defined as
\begin{eqnarray}
\omega_k = 
\left\{
\begin{array}{cccccc} 2\pi k / (N \Delta t) & , & k \le N/2 &  &  & \\
-2\pi k / (N \Delta t)  & , & k > N/2 &  &  &
\end{array}
\right.
\label{omega}
\end{eqnarray}

Using (\ref{wavelet_freq}) and a standard Fourier transform routine, 
one can calculate the continuous wavelet transform (for a
given s) at all n simultaneously.
Wavelets coefficients allow an estimation of signal power. 
The global wavelet power spectrum is defined as $|W_n(s)|^2$
\cite{Guide} and the total signal power $\bar{W}^2 $ can be expressed as follows:
\begin{equation}
\sum\limits_{j=0}^{M-1} |W_j|^2 = \frac{1}{N} \sum\limits_{j=0}^{M-1} \sum\limits_{k=0}^{N-1} |W_{k,j}|^2 = 
\frac{1}{N} \sum\limits_{j=0}^{M-1} \sum\limits_{k=0}^{N-1} |\bar{X}_k \times \bar{\Psi}_{k,j}|^2 
\label{wavelet_power}
\end{equation}

A sum of products of Fourier coefficients calculated in a 32-point 
Fast Fourier Transform (FFT32) routine 
for ADC data data ($x_n$) in each clock cycle with
pre-calculated Fourier coefficients of the reference wavelet 
gives an estimation of the power for selected types of the wavelet.
The on-line power analysis requires simultaneous 
calculations of the power for many (i.e. m) wavelets with the various 
frequency scale s. 

We are considering two-dimensional analysis of the spectrogram containing a distribution 
of partial power contributions $|W_k(s)|^2$ for the Fourier 
index k vs. m scaled wavelets.
The on-line analysis is
simplified and strongly depends on available FPGA resources.
Such an analysis of the radio signals in real time for high frequency
sampling is a challenge and it requires very efficient algorithm
implementation into the FPGA. 
Fortunately,  variable precision DSP blocks in the newest Altera Cyclone V
family offer high-performance, power-optimized, and fully registered multiplication operations
for 9-bit, 18-bit, and 27-bit word lengths.

\section{32-point FFT Algorithm}

Let us consider the N-point DFT $\bar{X}$  
\begin{equation}
\bar{X}_{k= 0,\ldots, N-1} = \sum\limits_{n=0}^{N-1} x_n e^{-2i\pi kn/N}
\label{FFT32eqn}
\end{equation}
where the $x_n$, as samples from an ADC chip, are real.
The formula  (\ref{FFT32eqn}) can be split into two or more parts by rearranging 
the sum and indices.
The standard approach for formula simplification is through a Radix-2 
Decimation-in-Time (DiT)
or Decimation-in-Frequency algorithm (DiF).

For Radix-2 DiT, we get :
\begin{equation}
\bar{X}_{k} = \sum\limits_{n=0}^{\frac{N}{2}-1} x_{2n} e^{-i \frac{2\pi kn}{N/2}}  
+ e^{-i \frac{2\pi k}{N}} \sum\limits_{n=0}^{\frac{N}{2}-1} x_{2n+1} 
e^{-i \frac{2\pi kn}{N/2} } =
\label{eqn_DiT_1}
\end{equation}
\[
= DFT_{\frac{N}{2}}[x_0,x_2,...,x_{N-2}] + W^k_N \times DFT _{\frac{N}{2}}[x_1,x_3,...,x_{N-1}]
\label{eqn_DiT_3}
\]

The N-point DFT can be easily split into two N/2-point transforms.
Outputs from DFT procedures are complex. So, a calculation of final DFT 
coefficients by using a DiT algorithm
requires complex multiplication for final merging of data from parallel DFT 
procedures with lower order, i.e., multiplication 
of twiddle factors $W_N^k =  e^{-i\frac{2\pi k}{N}}$.

Altera$\textsuperscript{\textregistered}$  
provides a library routine of a complex multiplication in the FPGA.
However, a 16x16 bit operation requires 6 DSP embedded 9x9 multipliers,
even in the most economical (canonical) mode. 
Generally, a complex multiplication in the FPGA is rather resource-intensive 
and if possible it should be replaced by a
multiplication by real variables.

For Radix-2 DiF, we get :
\begin{equation}
\bar{X}_{2k} = \sum\limits_{n=0}^{N-1} x_n W^{2kn}_N = 
DFT_{\frac{N}{2}}\left[x_n +x_{n+\frac{N}{2}}\right]
\label{eqn_DiF_1}
\end{equation}
\[
\bar{X}_{2k+1} = \sum\limits_{n=0}^{N-1} x_n W^{(2k+1)n}_N = 
DFT_{\frac{N}{2}}\left[\left(x_n - x_{n+\frac{N}{2}}\right)W^n_N\right]
\label{eqn_DiF_2}
\]

The standard Radix-2 DIF algorithm rearranges the
DFT equation (\ref{FFT32eqn}) into two parts: 
computation of the even-numbered discrete-frequency indices $\bar{X(k)}$
for k=[0,2,4,$\ldots$], and computation of the odd-numbered indices
k=[1,3,5,$\ldots$]. This corresponds to splitting an N-point DFT into
two k = N/2-point routines. The first corresponding twiddle factor is 
$e^{-i\frac{2\pi}{N}\frac{N}{2}}=-1$. 
The first operations are simple sums and subtractions of real variables. 
Each operation related to the consecutive
twiddle factor will be performed in a single clock cycle. 

The algorithm of Decimation in Frequency used for the 32-point DFT allows 
splitting eq. \ref{FFT32eqn} as follows:
\begin{equation}
\bar{X}_{k=2p} =  \sum\limits_{n=0}^{15} A_n e^{-i\pi kn/8} 
\hspace{5 mm} \Rightarrow \hspace{5 mm} FFT16_{even}
\label{FFT16_even}
\end{equation}
\begin{equation}
\bar{X}_{k=2p+1} =  \sum\limits_{n=0}^{15} A_{n+16} e^{-i\pi (2p+1)n/16}
\label{FFT16_odd}
\end{equation}
where for n = 0,1,...,15
\begin{equation}
A_n =  x_n + x_{n+16} \hspace{5 mm} A_{n+16} = x_n - x_{n+16} 
\label{A_var}
\end{equation}
The next twiddle factors are: 
\begin{eqnarray}
W_B  =  e^{-i\pi /2} = - i \hspace{7 mm} W_C  =  e^{-i\pi /4} = \gamma (1 - i)   \\
W_D  =  e^{-i\pi /8} =  \alpha - i \beta \hspace{3 mm}  W_E  =  e^{-i\pi /16} = \xi - i \eta\\
W_F  =  e^{-3i\pi /16} = \sigma - i\rho
\label{twiddle_factors}
\end{eqnarray}
where
\begin{eqnarray}
\gamma & = & cos(\pi/4)  \hspace{3 mm} \alpha  =  cos(\pi/8)  \hspace{3 mm} \beta = sin(\pi/8)\\
\xi & = & cos(\pi/16) \hspace{6 mm}  \eta = sin(\pi/16) \\
\sigma & = & cos(3 \pi/16) \hspace{4 mm}  \rho = sin(3 \pi/16)
\label{coeff}
\end{eqnarray}

The DiF algorithm and further optimizations take into account only 
FFT coefficients with indices 
k = 0,...,15. Due to real input data ($x_{0,...31}$), the higher FFT
coefficients have well known symmetry : $Re \bar{X}_{32-n} = 
Re \bar{X}_n$ and $Im \bar{X}_{32-n} = -Im \bar{X}_n$ ($n>0$).
Calculation of the $\bar{X}_{0,...15}$ according to the pure Radix-2 DiF 
algorithm requires 8 pipeline stages. For $\bar{X}_{0,4,8,12,16}$,
2 pipeline stages are necessary only for synchronization. 

According to eq. (\ref{FFT16_even}) all $\bar{X}_{0,2,4,...,14}$ 
with even indices could be calculated by the algorithm
presented in \cite{FFT16}. Variables $x_n$ in Fig. 2 in \cite{FFT16} 
would be replaced by $A_n$ according to eq. (\ref{A_var}).
An application of a modified algorithm reduces the number of $9\times9$ 
multipliers from 12 to 10 only, and shorten the pipeline chain on 
stages (the last 2 stages are simple registers for synchronization).

For the odd indices, stages $B$ and $C$ for 
k=16,...,19 and k = 24,...27 are pure delay lines, while for neighboring 
indices  k=20,...,23 and k = 28,...31, mathematical operations are performed 
in a cascade. Let us multiply the $A_{16,...19}$ and $A_{24,...27}$
by the factor $\lambda = {\gamma}^{-1}$. Then to adjust variables in 
the $C$ stage for odd FFT coefficients (for k = 20,21,22,33 and k = 28,29,30,31), we set
\begin{equation}
C_{k} = \lambda \times \gamma  =  B_{k}  
\label{C_B}
\end{equation}
So, by such a redefinition, the $C$ stage for odd FFT indices becomes a pure pipeline stage.
It can be removed with one pipeline stage for the even FFT indices.
In order to come back to the correct values, the coefficients in the $F$ stage can be simple redefined as:
\begin{eqnarray}
\alpha' & = &\gamma \times \alpha  \hspace{5 mm}  \beta' = \gamma \times \beta  \hspace{5 mm} \xi'  =  \gamma \times \xi  \\
\sigma' & = & \gamma \times \sigma  \hspace{6 mm}  \rho' = \gamma \times \rho  \hspace{6 mm}  \eta' = \gamma \times \eta
\label{redefinition}
\end{eqnarray}
but for indices k = 16,20,24,28 we have to use 4 additional multipliers. 
Nevertheless, at this cost we save one pipeline stage, and depending on 
the width of buses in the final FFT coefficients, we save at least 1000
logic elements.

%%%%%%%%%%%%%%%%%%%%%%%%%%%%%%%%%%%%%%%%%%%%%%%%%%%%%%%%%
%                                                                                                                                                       %
%                                                                          Figure 1                                                                %
%                                                                                                                                                       %
%%%%%%%%%%%%%%%%%%%%%%%%%%%%%%%%%%%%%%%%%%%%%%%%%%%%%%%%%
\begin{figure*}[!t]
\centering
\includegraphics[width=15cm,keepaspectratio]{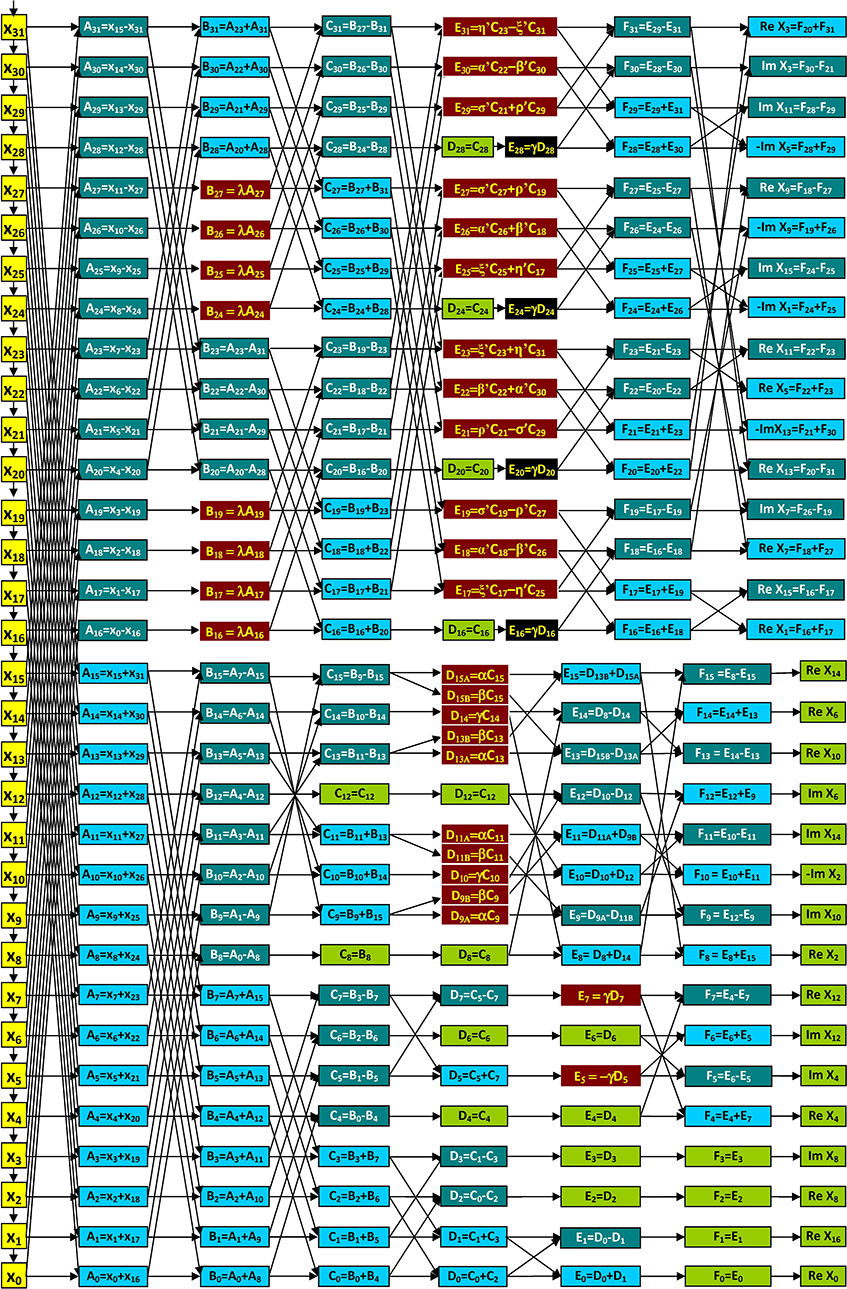}
\caption{An optimized structure with a reduced of a single pipeline stage 
at the cost of only 4 additional multipliers (8 DSP $9\times9$ blocks).}
\label{FFT32_opt}
\end{figure*}
%%%%%%%%%%%%%%%%%%%%%%%%%%%%%%%%%%%%%%%%%%%%%%%%%%%%%%%%%

\section{Wavelet Power Calculation}

Two families of reference Morlet wavelets were selected for the scaling factor 
$s^{-1} = \alpha$ = 0.04, 0.004, respectively:
\begin{equation}
f_{\alpha,freq}(k) = cos\left (2\pi k\frac{freq_{wavelet}}{sampling}\right ) exp(-\alpha \cdot k^2)
\label{wavelet_power_2}
\end{equation}
where -16$\le$k$\le$15, $sampling$ = 200 (MHz) \cite{ICRC2013}.

%%%%%%%%%%%%%%%%%%%%%%%%%%%%%%%%%%%%%%%%%%%%%%%%%%%%%%%%%
%                                                                                                                                                       %
%                                                                          Figure 2                                                                %
%                                                                                                                                                       %
%%%%%%%%%%%%%%%%%%%%%%%%%%%%%%%%%%%%%%%%%%%%%%%%%%%%%%%%%
\begin{figure}[h] 
\centering 
\includegraphics[width=\columnwidth,keepaspectratio] {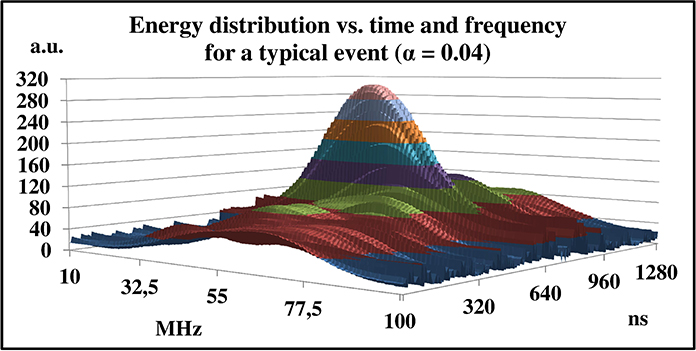}
\includegraphics[width=\columnwidth,keepaspectratio] {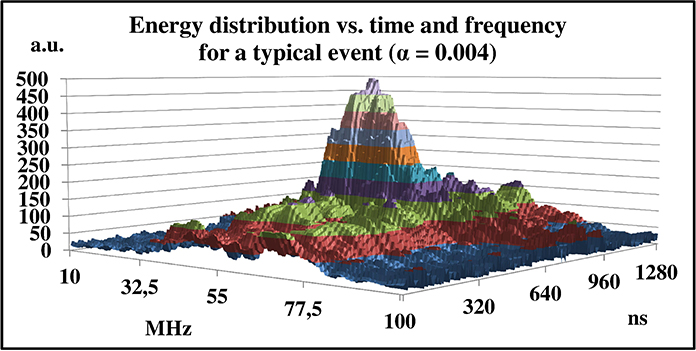}
\includegraphics[width=\columnwidth,keepaspectratio] {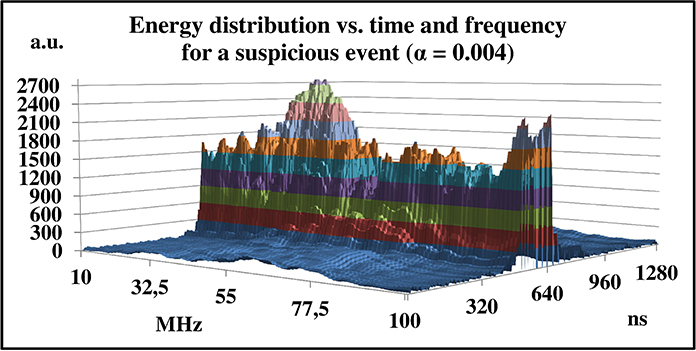}
\caption{An energy distribution as a function of time and frequency given for scaling factors
 $\alpha$ = 0.04 (1st graph) and $\alpha$ = 0.004 (2nd and 3rd) for 
typical events (1st and 2nd) as well as 
for a suspicious event (3rd graph). 
}
\label{events}
\end{figure}
%%%%%%%%%%%%%%%%%%%%%%%%%%%%%%%%%%%%%%%%%%%%%%%%%%%%%%%%%

Wavelets for $\alpha$ = 0.04 are squeezed to non-negligible values in a range of -10$\le$k$\le$10.
For  $\alpha$ = 0.004, wavelets are stretched to a larger range beyond  -16$\le$k$\le$15.  
$|\bar{X}_k|$  distributions are relatively wide for  $\alpha$ = 0.04; a separation between 
frequency ranges is limited. Much better separation is for  $\alpha$ = 0.004. This value 
reduces more a leakage between frequency bins. It was selected for preliminary estimation of the signal power, 
although it corresponds to longer stretch of time than the typical length of 150 ns and wavelets in the interval of 150 ns 
do not vanish completely as for $\alpha$ = 0.01. 

A sum of products of Fourier coefficients ($\bar{X}_k$), calculated in the 32-point 
Fast Fourier Transform (FFT32) routine 
for ADC data ($x_n$) (in each clock cycle), with
pre-calculated Fourier coefficients of the reference wavelet ($\bar{\Psi}_k$), 
gives an estimation of the power for selected types of the wavelet (Eq. \ref{wavelet_power}). 
The on-line power analysis requires simultaneous 
calculations of the power for many wavelets with the given scaling factor $s = \alpha^{-1}$. 
The data acquisition system is equipped in the band-pass filter with lower and higher cut-off frequencies 
at 30 and at 80 MHz, respectively.  Fig. \ref{events}a and Fig. \ref{events}b show an energy distribution for a typical radio event. 
The energy is cumulated mostly in a frequency range
of 40$\le$freq$\le$70 MHz. Fig. \ref{events}a  is calculated for the scaling factor $\alpha$ = 0.04 
and presents a rough structure of the energy distribution. 
This is due to an overlap of the wavelet Fourier distributions and spectral leakage between neighboring 
frequency sectors. For the scaling factor $\alpha$ = 0.004 the energy distribution is reconstructed with a very high precision (see Fig. \ref{events}b). 
%%%%%%%%%%%%%%%%%%%%%%%%%%%%%%%%%%%%%%%%%%%%%%%%%%%%%%%%%
%                                                                                                                                                       %
%                                                                          Figure 3                                                                %
%                                                                                                                                                       %
%%%%%%%%%%%%%%%%%%%%%%%%%%%%%%%%%%%%%%%%%%%%%%%%%%%%%%%%%
\begin{figure}[h] 
\centering 
\includegraphics[width=\columnwidth,keepaspectratio] {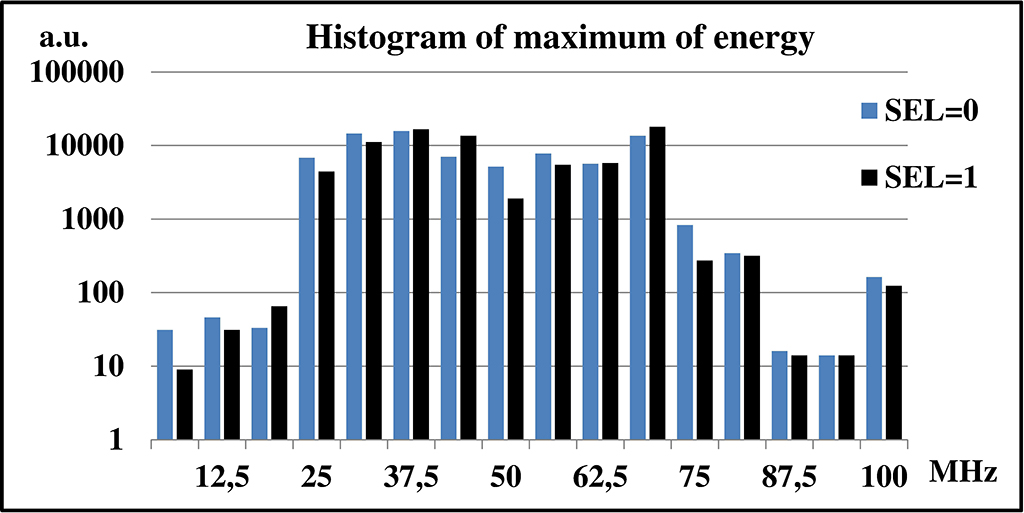}
\caption{A histogram of frequency indices when maximum of energy appears given for 77594 AERA events.
Energy distribution for each time bin of an event is scaled to get maximal value of 100\%. 
The histogram is built from time bins of events corresponding to a maximal signals. }
\label{AERA}
\end{figure}
%%%%%%%%%%%%%%%%%%%%%%%%%%%%%%%%%%%%%%%%%%%%%%%%%%%%%%%%%

The analysis of AERA data allowed an identification of  several "suspicious" events with significantly different spectral characteristics
(Fig. \ref{events}c) from "typical" ones (Fig. \ref{events}b). 
Fig. \ref{AERA} shows a distribution of energy vs. frequency. Energy is concentrated mostly in an expected frequency range 30 - 80 MHz
(in agreement with a characteristic of the band-pass filter).
As an estimator for a selection of untypical events we defined a spectral energy contributions 
(for each reference wavelet) averaged for 32 or 16 time bins
around the maximal signal. Next, the spectral distortion estimator (SDE) is calculated as follows:
\begin{equation}
SDE =  \sum\limits_{k=k_{5 MHz}}^{k_{low}-1} |W_k|^2 + \sum\limits_{k=k_{high}}^{k_{100 MHz}} |W_k|^2 
 -  \sum\limits_{k=k_{low}}^{k_{high}} |W_k|^2
\label{SDE}
\end{equation}

SDE $\ge$ 0 selects events with a significant energy contribution in peripheral energy ranges 
(traces marked as "suspicious").
Table \ref{rate} shows that the SDE rate for many widths of peripheral bands (calculated for 77594 AERA events). 
Conservatively calculated variants A and B show negligible amount of events. 
However, for more realistic conditions taking into account
borders of analog band-pass filter, the rate reaches few percents.

The energy contributions for the reference wavelets shown in Table \ref{rate}
are calculated as follows (denoted with SEL=1 or 0, respectively):
\begin{itemize}
\item{$ |W_j|^2 =   \sum\limits_{k=0}^{N-1} |\bar{X}_k \times \bar{\Psi}_{k,j}|^2 $ \qquad    SEL=1}
\item{$ |W_j|^2 =  |\bar{X}_k \times \bar{\Psi}_{k,k}|^2 $  \qquad \qquad    SEL=0}
\end{itemize}
with averaging over 32 or 16 samples (AVG=32 or 16) for two scaling factors
$\alpha$ = 0.004 and $\alpha$ = 0.0001, respectively.

For all configurations there is no any event corresponding to A conditions (sum of energy in
peripheral bands : 6.25-12.5 MHz + 97.75-100 MHz). However, for bands 
(6.25-31.25 MHz + 81.25-100 MHz) which should be cut-off
by the analog band-pass filter (variant D), few percents of "untypical" events were found.

%%%%%%%%%%%%%%%%%%%%%%%%%%%%%%%%%%%%%%%%%%%%%%%%%%%%%%%%%
%                                                                                                                                                       %
%                                                                    Table 1                                                                       %
%                                                                                                                                                       %
%%%%%%%%%%%%%%%%%%%%%%%%%%%%%%%%%%%%%%%%%%%%%%%%%%%%%%%%%
\begin{table}[b]
\centering\caption{\label{rate}A rate of "suspicious" for SDE $\ge$ 0}
\begin{tabular}{|c|c||c|c|c|c|c|}
\hline
low & SEL   & 6.25-12.5 & 6.25-18.75 & 6.25-25 & 6.25-31.25 \\
high &/AVG & 93.75-100 & 87.5-100 & 81.25-100 & 81.25-100  \\
\hline      
variant &  & A & B & C & D\\                                              
\hline                                                    
rate      &1-32  &0.000\% & 0.005\% & 0.14\% & 1.74\% \\
            &1-16  &0.000\% & 0.006\% & 0.15\% & 1.78\% \\
$\alpha$ =       &0-32  &0.000\% & 0.015\% & 0.22\% & 3.05\% \\
0.004   &0-16  &0.000\% & 0.019\% & 0.23\% & 3.11\% \\
\hline
rate      &1-32  &0.000\% & 0.005\% & 0.14\% & 1.78\% \\
         &1-16  &0.000\% & 0.009\% & 0.15\% & 1.80\% \\
$\alpha$=     &0-32  &0.000\% & 0.017\% & 0.23\% & 3.10\% \\
0.0001 &0-16  &0.000\% & 0.021\% & 0.24\% & 3.15\% \\
\hline
\end{tabular}
\end{table}
%%%%%%%%%%%%%%%%%%%%%%%%%%%%%%%%%%%%%%%%%%%%%%%%%%%%%%%%%

Table \ref{rate} shows that discrepancies a) by averaging on 32 or 16 samples as well as b) scaling factors
$\alpha$ = 0.004 or 0.0001, 
are negligible. Averaging on 16 samples simplifies the FPGA code.
By the scaling factor $\alpha$ = 0.0001 the spectral leakage between neighboring wavelet coefficients
is lower than for $\alpha$ = 0.004 and the energy contributions for the reference wavelets can be
calculated by a simple product $|\bar{X}_k| \times C_{k,k} $ (SEL=0, instead of the sum (SEL=1).
An elimination of sums significantly simplifies the FPGA code.

Number of events with maximal signal in frequency ranges [6.25-25 MHz] and [81.25-100 MHz]
is 1057 (1.36 \%). This gives the same level as for the variant D. We estimate that $\sim$1-3\% of
events have "untypical" spectral characteristics. Analyzed data are recorded directly from the ADC
(are unfiltered). Due to the analog band-pass filter events like in Fig. \ref{events}c, actually, should not
appear in a data stream. Their origin up to now is not clear. If they correspond to same rare
signal structures an on-line spectral analysis and spectral trigger may support a recognition of
some new phenomena. Modern FPGAs allows an implementation of more and more sophisticated
algorithms.

%%%%%%%%%%%%%%%%%%%%%%%%%%%%%%%%%%%%%%%%%%%%%%%%%%%%%%%%%
%                                                                                                                                                       %
%                                                                          Figure 4                                                                %
%                                                                                                                                                       %
%%%%%%%%%%%%%%%%%%%%%%%%%%%%%%%%%%%%%%%%%%%%%%%%%%%%%%%%%
\begin{figure}[t]
\centering
\includegraphics[width=\columnwidth,keepaspectratio] {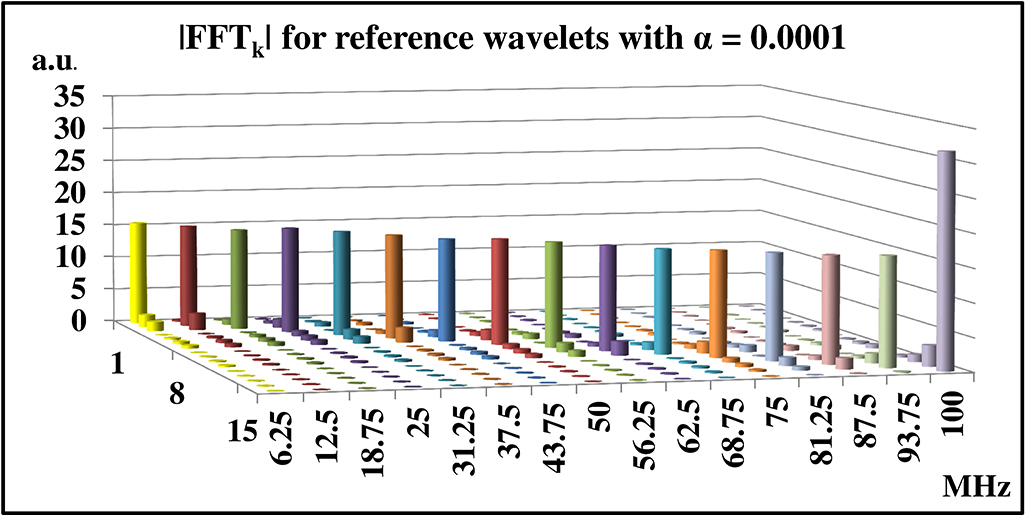}
\caption{The Fourier spectrum for 16 reference wavelets scaled for $\alpha = 0.0001$ }
\label{Morlet}
\end{figure}
%%%%%%%%%%%%%%%%%%%%%%%%%%%%%%%%%%%%%%%%%%%%%%%%%%%%%%%%%

\section{FPGA implementation}

The graphs in Fig. \ref{events} are generated for 255 channels (255 reference wavelets). 
This amount is certainly too large for an implementation in the FPGA. 
The elementary component from Eq. (\ref{wavelet_power}) is a product of two complex numbers: 
Fourier coefficients ($\bar{X}_k$)  of real data ($x_n$) and Fourier coefficients of the reference wavelet ($\bar{\Psi}_k$). 
The simplest way to calculate this product on-line in the FPGA is usage the 
Altera$\textsuperscript{\textregistered}$  IP routine ALTMULT\_COMPLEX. 
However,  the wavelet spectral coefficients are predefined  as constants for arbitrarily selected reference wavelets.
A module of a complex product  
$|\bar{X}_k \cdot \bar{\Psi}_{k,j}|^2$ 
can be calculated as a real product of
$|\bar{X}_k|^2 \cdot |\bar{\Psi}_{k,j}|^2 =  |\bar{X}_k|^2 \cdot C_{k,j}$ (where $C_{k,j}$ are constants - 
previously pre-calculated wavelet coefficients). We need calculate on-line only $|\bar{X}|^2$ and next
multiply by constant factors.

Fig. \ref{Morlet} shows the Fourier spectrum for only 16 reference wavelets scaled for a possible big resolution
($\alpha = 0.0001$). 
The power contribution to the sum in Eq. \ref{wavelet_power} comes in fact from a single index - k.
The contribution from neighboring k-1 and k+1 indices are less than 2-3 \%.
Thus, the wavelet engine can use much less DSP blocks focusing on only a single index which gives
a fundamental contribution to the total power. Practically, it means that 
$|W_k|^2 = |X_k|^2$ for all indices except the maximal one, which should be scaled by the factor of
23,74611/11,73 $\sim$ 2. A multiplication by a factor of 2 in the FPGA is accomplished by
a connection of the bus to the next pipeline stage $S^{stage+1}[BUS..1] = S^{stage}[BUS-1..0]$.

AERA uses 14-bit ADCs. In the algorithm from Fig.\ref{FFT32_opt} each next pipeline stage 
extends a width of data bus on a single bit. Outputs of $\bar{X}_{Re}$ and $\bar{X}_{Im}$ have
20-bit buses. This provides a negligible approximation errors. A maximal error of
8 ADC-counts by 20-bit resolution corresponds to the error of 0.008\%. However, a square of
20-bit variable would require 4 DSP blocks. A reduction of a bus width to 16-bit only decreases the DSP blocks
utilization to 2 DSP per module, however at the cost of an accuracy. Differences between exact
and FPGA calculated modules reaches 1.5-2\% but only for less than 1\% of events.

\section{Trigger}

Spectral energy contributions $|W_{k,j}|^2$ (Eq. \ref{wavelet_power}) are integrated by
boxcar \cite{IEEE1999} averaging 16 samples to get $|W_k|^2$ (Eq. \ref{SDE}). Results from
Table \ref{rate} justify a shorter boxcar chain for FPGA code simplification and resources economy.

If the MSB (sign bit) of the SDE equals to 0 it means the analyzing events has an untypical spectral
characteristic. Trigger is generated. Data is continuously written into the left port of the dual-port RAM.
In order to catch the profile of the event, trigger is delayed in RAM-based shift register on 512 clock
cycles. Thus data is frozen in 1024-word DPRAM. The profile is next read by the NIOS processor and
sent via UART to the external host (PC).

\section{IIR-notch filter}

The main purpose of the IIR (Infinite Impulse Response) notch filter \cite{KELLEY} in the radio detector is 
to increase the signal-to-noise ratio 
before triggering. The previously implemented filter based on the FFT approach \cite{FFT_AERA} 
was power consuming and has been replaced by the IIR-notch one. 
The IIR-filters operate on the time-domain and the output of the filter ($y_i$) is a linear combination of 
the input samples ($x_j$) and the delayed feedback output samples ($y_j$): 
\begin{eqnarray}
y_i = x_i & - & (2\cos \omega_N\cdot x_{i-1})  + x_{i-2} +   \nonumber \\
                 & + & (2r\cos \omega_N\cdot y_{i-1}) - (r^2\cdot y_{i-2})  
\label{eq:notch}
\end{eqnarray}
The value of $r$ is to be between 0 and 1, where a value close to 1 indicates a very narrow transmitter. 
A typical value chosen is $r = 0.99$.
$\omega_N$ is the normalized filter frequency, which is determined by the notch frequency $f_N$ and the sampling
 frequency $f_S$ as $\omega_N = (2\pi f_N)/f_S$.

%%%%%%%%%%%%%%%%%%%%%%%%%%%%%%%%%%%%%%%%%%%%%%%%%%%%%%%%%
%                                                                                                                                                       %
%                                                                          Figure 5                                                                %
%                                                                                                                                                       %
%%%%%%%%%%%%%%%%%%%%%%%%%%%%%%%%%%%%%%%%%%%%%%%%%%%%%%%%%
\begin{figure}[ht]
\centering
\includegraphics[width=\columnwidth,keepaspectratio] {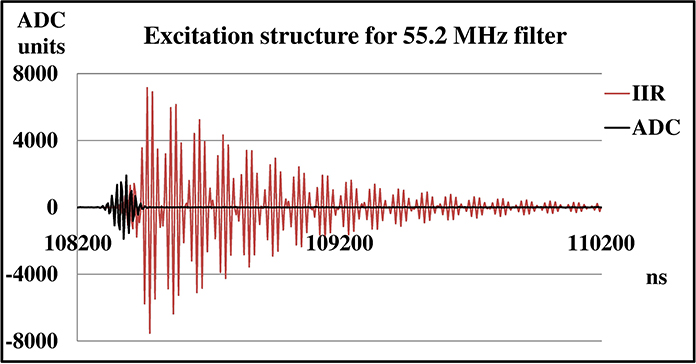}
\caption{Excitations appearing in two different IIR-notch filters for test frequencies 55.2 MHz and 25.39 MHz, respectively.
The upper graph shows a zoom of excitation for 55.20 MHz filter. The bottom graph 
shows a relatively often excitations for 25.39 MHz of the base frequency. Let us notice that excitations 
have huge amplitude, 4 timer larger than input signals.}
\label{excitations}
\end{figure}
%%%%%%%%%%%%%%%%%%%%%%%%%%%%%%%%%%%%%%%%%%%%%%%%%%%%%%%%%

Generally, the IIR filter may contain unstable regions, due to internal feedbacks. 
The IIR-filter characteristics are optimized to cut-off narrow bands corresponding to radio transmitters operating 
in region of radio detectors. Band reject regions should be narrow enough in order not to affect real cosmic ray signals.
On the other hand very narrow filters are liable to potential instabilities.
Radio stations contain 4-stage filters implemented in a cascade way suppressing 4 narrow bands. 
We  have simulated responses of the filter for
the following band-reject frequencies: 27.12, 40.9, 55.2 and 70.7 MHz, respectively, with r = 0.99. 
Unfortunately, we found in Quartus simulations  instabilities for some configurations. 
Fig. \ref{excitations}a shows an excitation on the output of the 3rd filter (55.2 MHz cut-off) 
for signal packages driving the filter cascade. For 55.2 MHz filter excitations appear relatively rarely, 
however, if occurs it may generate a spurious trigger.
For others frequencies we observed much higher rate (Fig. \ref{excitations}b).

%%%%%%%%%%%%%%%%%%%%%%%%%%%%%%%%%%%%%%%%%%%%%%%%%%%%%%%%%
%                                                                                                                                                       %
%                                                                          Figure 6                                                                %
%                                                                                                                                                       %
%%%%%%%%%%%%%%%%%%%%%%%%%%%%%%%%%%%%%%%%%%%%%%%%%%%%%%%%%
\begin{figure}[t]
\centering
\includegraphics[width=\columnwidth,keepaspectratio] {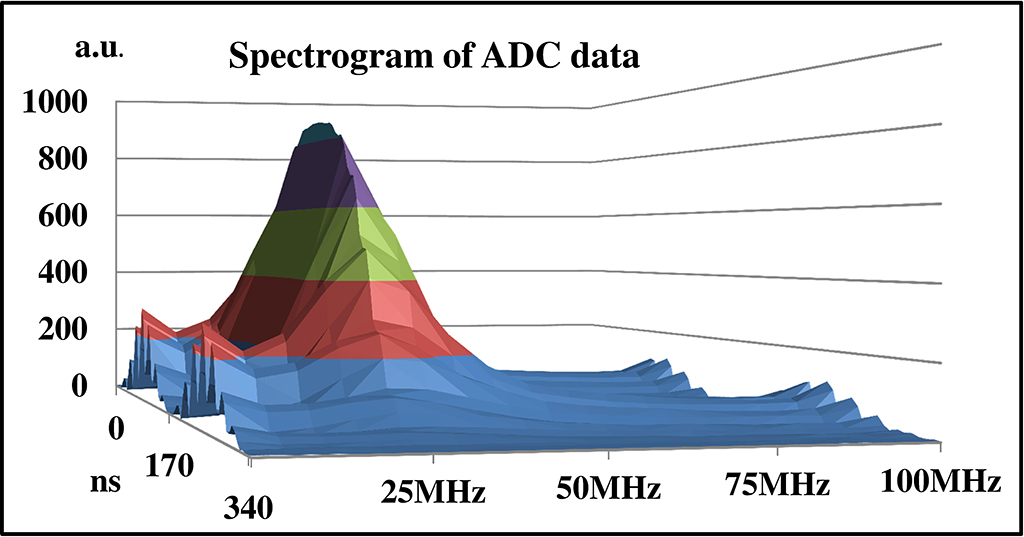}
\includegraphics[width=\columnwidth,keepaspectratio] {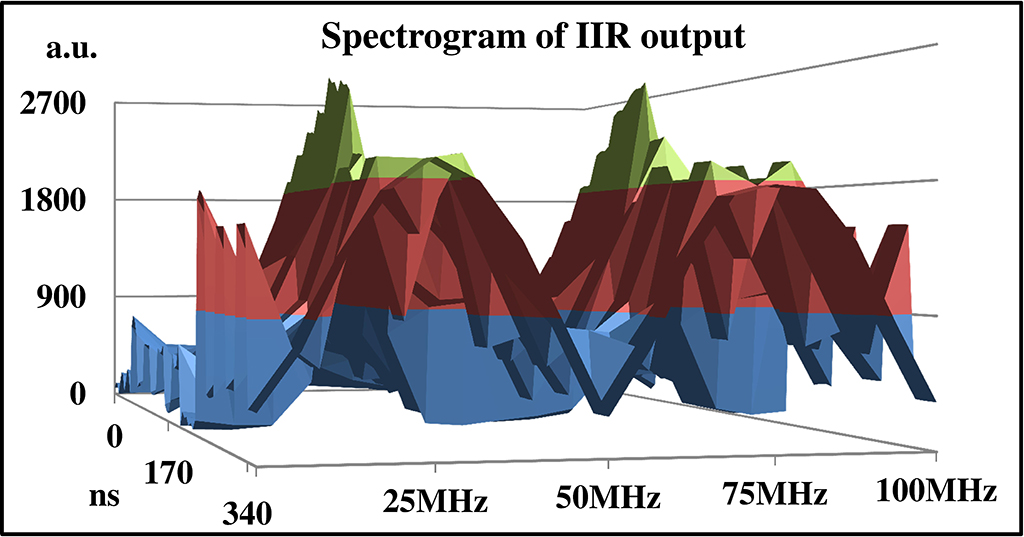}
\caption{Spectrograms of ADC test input (top) and IIR output (bottom).  }
\label{25_39MHz}
\end{figure}
%%%%%%%%%%%%%%%%%%%%%%%%%%%%%%%%%%%%%%%%%%%%%%%%%%%%%%%%%

%%%%%%%%%%%%%%%%%%%%%%%%%%%%%%%%%%%%%%%%%%%%%%%%%%%%%%%%%
%                                                                                                                                                       %
%                                                                    Table 2                                                                       %
%                                                                                                                                                       %
%%%%%%%%%%%%%%%%%%%%%%%%%%%%%%%%%%%%%%%%%%%%%%%%%%%%%%%%%
\begin{table}[b]
\centering\caption{\label{exc} Excitations evidence vs. compiler parameters}
\begin{tabular}{|c||c|c|c|c|c|c|}
\hline
Effort  & compiler & version &system & $f_{max}$ & exc.&exc.\\
level  &                     &                  &               &                       & sim & mea\\
\hline \hline                                                    
Normal & 9.1sp2 & Web free & XP 32-Bit& 232.94 & YES& NO \\
Extra     & 9.1sp2 & Web free & XP 32-Bit& 236.69 & NO & NO\\ \hline
Normal & 9.1        & Full paid & XP 32-Bit & 232.94 & YES & NO\\
Extra     & 9.1        & Full paid & XP 32-Bit & 236.69 & NO & NO\\ \hline
Normal & 9.1         & Full paid & W7 64-Bit & 236.69 & NO & NO\\ 
Extra    & 9.1         & Full paid & W7 64-Bit & 237.02 & NO & NO\\ \hline
\end{tabular}
\end{table}
%%%%%%%%%%%%%%%%%%%%%%%%%%%%%%%%%%%%%%%%%%%%%%%%%%%%%%%%%

A wavelet analysis of the IIR data output shows a dramatic change of spectral characteristic for signals with excitations. 
A power concentrated originally close to 25 MHz (Fig. \ref{25_39MHz}a) spreads over much higher frequency range
(Fig. \ref{25_39MHz}b). Spectral characteristics of excited events (Fig. \ref{25_39MHz}b) are similar to "suspicious" real events
(Fig. \ref{events}c).
At first we suspected that "suspicious" events in AERA database are registered by spurious triggers generated by
IIR-notch filter for disadvantageous correlations between the filter cut-off and a basic signal frequency.
But a careful analysis showed that analyzed events came only from unfiltered ADC traces. The "suspicious" event
could not be generated by a filter. So, the origin of the strange spectral characteristics remained still mysterious.

Next steps showed that excitations depend on switches in the compiler and its version!
Originally, excitations were found in simulations for Quartus 9.1 sp2 free version on the XP 32-Bit OS.
Simulation for other versions or systems did not confirm an evidence of any excitations. 
Finally, we found that excitations appear for "Normal" position in Effort level (Setting option) for
32-Bit compilers working on XP OS. 64-Bit version running on Windows 7 (64-Bit) does not generate
spurious filter responses (Table \ref{exc}). Nevertheless, the registered performance for "strange" configurations is high enough
(a safety margin is larger than 10\%) to provide a stable internal data propagations without any potential
hazards or races.

Fortunately, an evidence of excitations has not been confirmed in the laboratory tests for on the Cyclone III FPGA with 
sof generated by the Quartus 9.1 sp2 and 9.1 versions, independently free or paid.
Problems suggested by Quartus simulations did not occur in reality
(Table \ref{exc}).

\section{Laboratory tests}

\subsection{Excitations}

As sources of artificial events we used:
\begin{enumerate}
\item{ the IIR-notch filter implemented in FPGAs of two Altera development kits:
\begin{enumerate}
\item{with Cyclone III EP3C120F780C7 supported by ICB-HSMC daughter card with RS232 serial port allowing
a connection between the NIOS and the PC, Cyclone III was programmed by the sof file generated by Quartus 9.1 sp2 on 32-bit XP OS,}
\item{with Cyclone V E 5CEFA7F31I7 supported by 
Altera HSMC-ADC-BRIDGE and Texas Instr. ADS4249EVM (Evaluation Module - with double channel 
14-bit ADC with max. 250 MHz sampling), Cyclone V was programmed by the sof file generated by the licensed version
of Quartus 13.1 on Windows7 (64-bit OS)}
\end{enumerate}
\item{arbitrary pattern generator Tektronix AFG3252C driving the chain b) }
}
\end{enumerate}

When the wavelet engine recognized a signal with untypical spectral characteristic a generated trigger 
frozen 1024 samples of averaged $\bar{X}_{k=1,...15}$ in Dual-Port RAM embedded memory of 1024 words 
which next were transmitted to the PC by a NIOS processor via RS232 serial port.
Amplitudes of input test vectors were lower than 2048 ADC-units. 
Cyclone III was driven by the test vector stored in the embedded ROM, Cyclone V was driven from 
the arbitrary pattern generator Tektronix AFG3252C (pattern of 128 kwords of 14-bit data or burst signals)
via ADS4249EVM and HSMC-ADC-BRIDGE.
Cyclone V supports On Chip Termination (OCT), Cyclone III does not. The Cyclone III development kit was not equipped 
in termination resistors. Trigger was set:
\begin{itemize}
\item{just after the notch filter recognizing possible "sparks" with an amplitude much higher than maximal input signal - 
excitations from Fig. \ref{excitations}b), }
\item{after the wavelet engine when SDE output indicated a negative values.}
\end{itemize}

Fortunately,  we did not find excitations in Cyclone III test setup with sof from Quartus 9.1 (also sp2).
Spurious events were not found both by amplitude comparator just after the notch filter as well as by the wavelet trigger
analyzing an energy contribution in frequency domain.
According expectations, no any spurious events were found in the Cyclone V test platform.
Selected patterns of real AERA events (with untypical spectral characteristics) driving the FPGA from the arbitrary pattern 
generator AFG3252C
generated expected rates of wavelet trigger. Trigger rate for patterns from AFG3252C wired OR with noise from
33250A remains on the same level in a relatively wide range of additional noise.   

%%%%%%%%%%%%%%%%%%%%%%%%%%%%%%%%%%%%%%%%%%%%%%%%%%%%%%%%%
%                                                                                                                                                       %
%                                                                          Figure 12                                                              %
%                                                                                                                                                       %
%%%%%%%%%%%%%%%%%%%%%%%%%%%%%%%%%%%%%%%%%%%%%%%%%%%%%%%%%
\begin{figure}[h]
\centering
\includegraphics[width=\columnwidth,keepaspectratio] {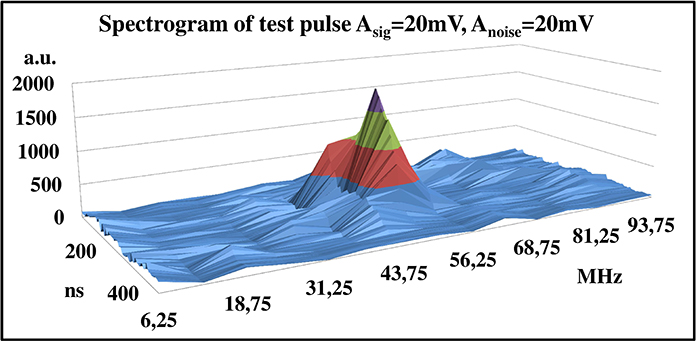}
\caption{Spectral characteristics measured a test pulse on Cyclone V development kit driven from AFG3252C
with an additional noise from 33250A.}
\label{test_mea}
\end{figure}
%%%%%%%%%%%%%%%%%%%%%%%%%%%%%%%%%%%%%%%%%%%%%%%%%%%%%%%%%

\subsection{Spectral characteristics}

We tested also an influence of the noise on spectral characteristics short pulses typical for
radio events (50.62 MHz sine wave enveloped by a Gauss bell). 
Even for a noise level comparable with a level of signal (Fig. \ref{test_mea}a) the signal
spectral contribution is clearly visible and easy to detect. However, for 3 times higher noise
(Fig. \ref{test_mea}b) we observe a dramatic change of spectra practically making impossible
a signal recognition. Let us see that for higher signals (with the same signal to noise ratio - Fig. \ref{test_mea}c) 
contributions for very low and very high frequencies are lower in comparison to relatively low signals, 
when quantization of signals additionally affect the spectra.

Practically, the input amplifiers seem to be configured with higher amplification factors agreeing on relatively
high (but white) noise than to use low level quantized ADC signal significantly affected by digitization processes.
It means, the amplitude of the signal should not be too low, a contributed noise reduces a errors coming from quantization 
processes.

The AHDL code implemented into Cyclone V 5CEFA7F31I7 FPGA has high enough register performance to
provide 15\% safety margin (with 200 MHz sampling). 

\section{Conclusions}

An evidence of untypical spectral characteristic in real AERA events was a motivation to develop
a FPGA data processing block allowing on-line a recognition of non-standard events.
The origin of these events maybe "natural" (real new phenomena) 
or "apparatus" (i.e. oscillations in filters or differential nonlinearities of ADCs).
Such a tool for sure helps to reject spurious triggers and, as more important, may help
to detect new category of events or phenomena.

Laboratory tests confirmed in real time data processing 
\begin{itemize}
\item{an efficiency of the wavelet trigger  
recognizing event with untypical spectral characteristics,}
\item{no instabilities in IIR-notch filters for sof files generated by all tested versions of Quartus compilers.}  
\end{itemize}

The presented above FPGA block can support standard triggers allowing a rejection of suspicious 
or too noisy events and a suppression of a database size.

\section*{Acknowledgment}
\vspace*{0.5cm}
\footnotesize{{\bf Acknowledgment:}
{This work is being developed for the next generation of cosmic ray detectors
supported by  the ASPERA-2 consortium 
and was funded by the Polish National Center of 
Researches and Development under NCBiR Grant No. ERA/NET/ASPERA/02/11
and by the National Science Centre (Poland) under NCN Grant No. 2013/08/M/ST9/00322

The author would like to thank Charles Timmermans from Radboud University Nijmegen, the Netherlands,
for files with AERA radio events.}}

\ifCLASSOPTIONcaptionsoff
  \newpage
\fi


\begin{thebibliography}{1}

%1
\bibitem{Alan}
H. R. Allan, ``Radio emission from extensive air showers'', 
\emph{Prog. in Elem. Part. and Cos. Ray Phys.}, vol. 10, pp. 171, 1971.

%2
\bibitem{Falcke_Gorham}
H. Falcke, P.W. Gorham, 
``Detecting radio emission from cosmic ray air showers and neutrinos 
with a digital radio telescope'',\emph{ Astropart. Phys.} vol. 19,
pp. 477-494, July 2003.

%3
\bibitem{Falcke} T. Huege, H. Falcke, Radio emission from cosmic ray air showers: Simulation results and parametrization, 
\emph{Astropart. Phys.} vol. 24, Issues 1-2, pp. 116–136, Sept. 2005.

%4
\bibitem{Huege} T. Huege, R. Urlich, R. Engel, Monte Carlo simulations of 
geo-synchrotron radio emission from CORSIKA-simulated air showers, 
\emph{Astropart. Phys.}  vol. 27, Issue 5, pp. 392-405, June 2007.

%5
\bibitem{Petrovic} J. Petrovic et. al, Radio emission of highly inclined cosmic ray air showers measured with LOPES, 
\emph{Astronomy \& Astrophys.},  462, pp. 389-395, 2007.

%6
\bibitem{LOPES} The LOPES Collaboration, Progress in air shower radio measurements: Detection of distant events, 
\emph{Astropart. Phys.} vol. 26, Issues 4-5, pp. 332-340, Dec. 2006.

%7
\bibitem{Daubechies} I. Daubechies, The wavelet transform time-frequency localization and signal analysis, 
\emph{IEEE Trans. Inform. Theory}, vol. 36, pp. 961–1004, 1990.

%8
\bibitem{Guide} C. Torrence, G.P. Compo, A Practical
Guide to Wavelet Analysis, \emph{Bulletin of the American Meteorological
Society}, 79, pp. 61-78, 1998.

%9
\bibitem{FFT16} Z. Szadkowski, 16-point discrete Fourier transform based on the Radix-2 FFT
algorithm implemented into cyclone FPGA as the UHECR trigger
for horizontal air showers in the Pierre Auger Observatory, 
\emph{Nucl. Instr. Meth.  A}, vol. 560, Issue 2, pp. 309-316, May 2006,

%10
\bibitem{ICRC2013}
Z. Szadkowski, FPGA Based Wavelet Trigger in Radio Detection of Cosmic Rays, \emph{Brazilian
Journal of Physics}, 2014 (in publication),

%11
\bibitem{IEEE1999} H. Gemmeke, A. Grindler, H. Keim, M. Kleifges, N. Kunka, Z. Szadkowski, D. Tcherniakhovski,
Design of the Trigger System for the Auger Fluorescence Detector,
\emph{IEEE Trans. on Nucl. Science}, vol. 47, Issue: 2, pp. 371-375 Part: 1, Apr. 2000.

%12
\bibitem{KELLEY}  J. L. Kelley, Data acquisition, triggering, and filtering at the Auger Engineering Radio Array, 
\emph{Nucl. Instr. Meth.  A}, vol. 725, pp. 133-136, Oct. 2013,

%13
\bibitem{FFT_AERA} A. Schmidt, H. Gemmeke, A. Haungs, K-H. Kampert, C. R\"uhle, Z. Szadkowski, 
FPGA Based Signal-Processing for Radio Detection of Cosmic Rays
\emph{IEEE Trans. on Nucl. Science}, vol. 58, pp. 1621-1627, Aug. 2011









\end{thebibliography}
\end{document}